# Animation of 3D Human Model Using Markerless Motion Capture Applied To Sports

Ashish Shingade[1] and Archana Ghotkar[2]

[1,2]Department of Computer Engineering, Pune Institute of Computer Technology, Pune, India

*ABSTRACT*

*Markerless motion capture is an active research in 3D virtualization. In proposed work we presented a system for markerless motion capture for 3D human character animation, paper presents a survey on motion and skeleton tracking techniques which are developed or are under development. The paper proposed a method to transform the motion of a performer to a 3D human character (model), the 3D human character performs similar movements as that of a performer in real time. In the proposed work, human model data will be captured by Kinect camera, processed data will be applied on 3D human model for animation. 3D human model is created using open source software (MakeHuman). Anticipated dataset for sport activity is considered as input which can be applied to any HCI application.*

## 1. INTRODUCTION

Motion capture and computer animation techniques have made significant progress in game and film industry. Detecting movements of people in 3D and displaying it in a 3D virtual scene is a research problem.

There are two ways for motion capture, marker based motion capture and markerless motion capture. The Marker based motion capture has many drawbacks, the major drawback is that the performer has to wear a suit which consists of sensor or markers on it and the process consist of handling multiple cameras placed in a room, hence markerless motion capture has become a major area of research. In markerless motion capture the performer doesn't have to wear a suit, but still markerless motion capture is a challenging task.

The markerless motion capture is not an easy task to perform as it requires extra effort, so enough good results cannot be obtained using a single ordinary camera, the process still requires a set of multiples cameras placed all over the room, which also increases cost of the overall system. With the development of depth cameras such as Microsoft Kinect has eased the task of motion capture, without requiring the burden of multiple cameras, hence it decreases the cost of overall system.
This paper gives a survey on various available techniques related to motion capture. All these techniques intend to develop an automated body motion capture technique which helps to create a digital animation in 3D, which can ease the task of animators. This paper also presents a discussion on depth camera and libraries that can be applied for skeleton tracking. The paper proposes markerless motion capture system for 3D human character animation which can be applied for any HCI application like gaming, film industry, motion analysis in sports and many more.

For proposed idea kinect camera has been used, which captures real time videos and gives output as a skeleton. We apply kinect with Microsoft kinect SDK for better performance, for creating a human model open source software of Make Human [29] is used and rigging is performed using algorithm mentioned in [22].

The paper is organized as follows. Section II gives the review of motion and skeleton tracking techniques, section III gives discussion on depth cameras and libraries that can be applied for skeleton tracking, section IV presents the methodology, mathematical model in section V and conclusion is presented in section VI.

## 2. REVIEW OF MOTION AND SKELETON TRACKING TECHNIQUES

The researchers have surveyed various approaches for body motion and skeleton tracking for various applications. The body motion and skeleton tracking techniques using an ordinary camera are not easy and require extensive time in developing. The survey of motion capture and motion capture for animation using kinect is presented in Table 1. And the survey of body and skeleton tracking techniques are presented by the technique used along with advantages, disadvantages and illustration as shown in Table 2.

Table 1. Surveys of different motion capture techniques using kinect camera.

| Authors | Application | Pros and Cons | Illustrations |
|---|---|---|---|
| Kairat Aitpayev, Jaafar Gaber [1] | Adding collision object for human body in augmented reality using kinect | 1. Easy to implement.<br>2. System not accurate enough.<br>3. Problems with measurement of bones. | 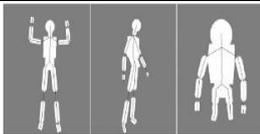 |
| Xiaolong Tong, Pin Xu, Xing Yan [2] | Skeleton animation motion data based on kinect | 1. Creation of standard motion data files in real time.<br>2. Reduces funding of implementation.<br>3. Jitter present in data achieved for foot.<br>4. Lack in Optimization of motion data. | 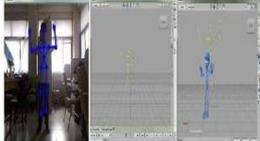 |
| Ming Zeng, Zhengcun Liu, Qinghao Meng, Zhengbiao Bai, Haiyan Jia [3] | Motion capture and reconstruction based on depth information using kinect | 1. Fairly accurate results obtained for real time 3D human body movements.<br>2. Good fidelity and low latency of system.<br>3. No support for occlusion handling. | 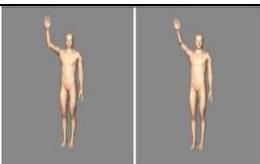 |
| Mian May, Feng Xu, Yebin Liu [4] | Animation of 3D characters form single depth camera | 1. Noise and errors with joints position are removed.<br>2. Due to removal of noise good results are obtained.<br>3. The deformation models pose is not that similar to the captured character.<br>4. Skinning is not done properly. | 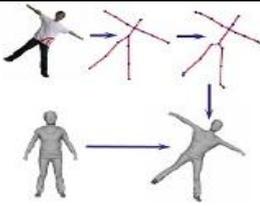 |

| Charles E. Colvin, Janie H. Babcock, John H. Forrest, Chase M. Stuart, Matthew J. Tonnemacher, and Wen-Shin Wang [5] | Multiple user motion capture and system engineering | 1. Support for mapping hand gestures. 2. Reduces funding of implementation. 3. Arm gestures not supported. | 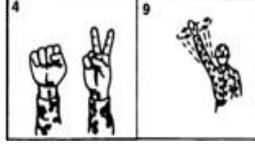 |
|---|---|---|---|
| Lucía Vera, Jesús Gimeno, Inmaculada Coma, and Marcos Fernández [6] | Augmented Mirror: Interactive augmented reality system based on kinect | 1. Head orientation, lip movements, facial expressions and automatic gestures (blink, hands, feet, etc.) are handled. 2. Occlusion is handled. 3. Finger tracking not supported. 4. Use of too many devices makes system difficult to implement. | 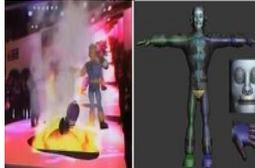 |
| Jing Tong, Jin Zhou, Ligang Liu, Zhigeng Pan, and Hao Yan [17] | Scanning 3D full human bodies using kinect | 1. Inference phenomenon is handled using multiple kinect. 2. Complex occlusions are handled. 2. Reduces funding of implementation. 3. Algorithm is memory efficient. 4. The quality of the reconstructed model is still poor. 5. Misalignments are still occurred 6. Unnatural bending in arm areas. | 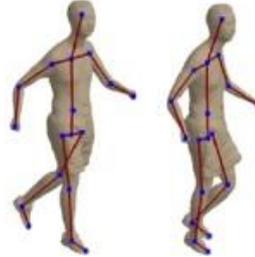 |
| Chanjira Sinthanayothin, Nonlapas Wongwaen, Wisarut Bholsithi [18] | Skeleton tracking using kinect sensor and displaying in 3D virtual scene | 1. Bone joint movements are detected in real time with correct position tracking. 2. No support for occlusion handling. | 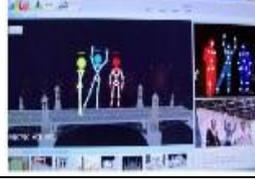 |
| Karina Hadad de Souza, Rosilane Ribeiro da Mota [19] | Motion Capture by Kinect | 1. Multiple kinect support for motion capture. 2. Increase in precision of system. 3. Occlusion handled with use of multiple kinect. 4. Not enough good performance. 5. With use of multiple kinect data processing increase. | 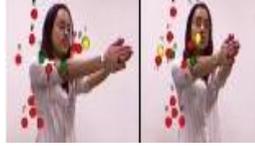 |
| Quanshi Zhang, XuanSong, Xiaowei Shao, Ryosuke Shibasaki, Huijing Zhao [20] | Unsupervised Skeleton Extraction and Motion Capture from Kinect Video via 3D Deformable Matching | 1. Approach is more robust than the traditional video-based and stereo-based approaches. 2. Good performance is obtained. 3. No support for occlusion in case when if a person folds his hands together. | 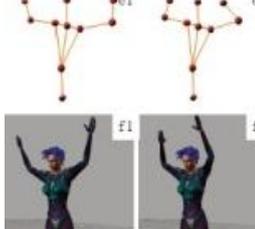 |

| Hubert Shum, Edmond S.L. Ho [21] | Real-time Physical Modeling of Character Movements with Microsoft Kinect | | 1. Proposed algorithm is computationally efficient and can be applied to a wide variety of interactive virtual reality applications<br>2. No support for occlusions and noises handling. | 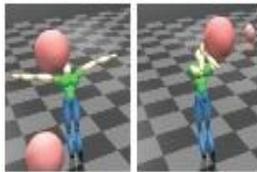 |
|---|---|---|---|---|

Table 2. Survey of body and skeleton tracking techniques.

| Authors | Applications | Techniques | Pros and Cons | Illustrations |
|---|---|---|---|---|
| Adso Fernandez-Baena, Antonio Susın, Xavier Lligadas [7] | Biomechanical validation of upper-body and lower-body joint movements of kinect motion capture data for rehabilitation treatments | Optical Motion Capture | 1. Reduces funding of implementation.<br>2. Comparison of Kinect motion capture with optical motion capture.<br>3. Fairly good results are obtained.<br>4. Lack of precision in system.<br>5. Approximation of joints and bones not done properly. | 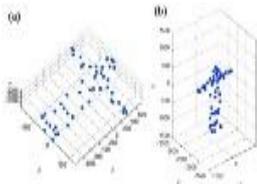 |
| Jamie Shotton, Andrew Fitzgibbon, Mat Cook, Toby Sharp, Mark Finocchio, Richard Moore, Alex Kipman, Andrew Blake [8] | Real time human pose recognition in parts from single depth camera | Randomized decision forests | 1. Quickly and accurately predicts 3D positions of body joints from single depth image, using no temporal information.<br>2. Ability to run the classifier in parallel on each pixel on a GPU to increase the speed.<br>3. Using large and highly varied training dataset to estimate body parts invariant to pose, body shape, clothing, etc. to pose the relation between two adjacent parts. | 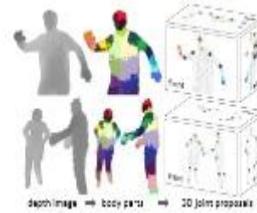 |
| Wei Qu, Dan Schonfeld [9] | Real-time decentralized articulated motion analysis and object tracking from videos | Decentralized articulated object tracking (DAOT), hierarchical articulated object tracking (HAOT). | 1. Fast and easy to implement.<br>2. Results not shown in case of self-occlusion due to the fact that it cannot handle pose relation between two adjacent parts. | 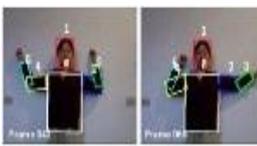 |
| Jesus Martınez del Rincon, Dimitrios Makris, Carlos Orrite, Jean-Christophe Nebel [10] | Tracking human position and lower body parts using Kalman and particle filters constrained by human | Position Tracking Based on a Kalman Filter, multiple-Particle-Filter Tracking Based on Two-Dimensional | 1. Bipedal motion is handled without any constraints.<br>2. Occlusion is seen case of pivot joints. | 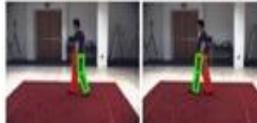 |

International Journal of Computer Graphics & Animation (IJCGA) Vol.4, No.1, January 2014

| | | | | |
|---|---|---|---|---|
| | biomechanics | Articulated Model | | |
| Raskin, Leonid, Michael Rudzsky, and Ehud Rivlin [11] | Gaussian process annealed particle filter for tracking and classification of articulated body motions | Gaussian Process Annealed Particle Filter | 1. Robust than hierarchical annealed particle filter. 2. Less errors. 4. In case of hugging motion classification fails. 5. Cross validation is needed to classify ambiguous types of motion. | 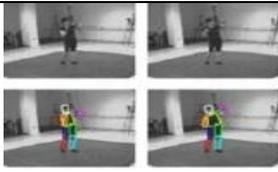 |
| Bernier, Olivier, Pascal Cheung-Mon-Chan, and Arnaud Bouguet [12] | Fast nonparametric belief propagation for real-time stereo articulated body tracking | Recursive Bayesian Tracking for Articulated Objects | 1. Good results shown on arm movements to various human tracking positions 2. Slow processing rate. 3. In case of complex occlusion system fails. | 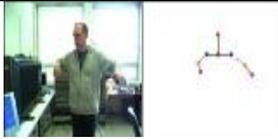 |
| Zhu, Rong, and Zhaoying Zhou [13] | A real-time articulated human motion tracking using tri-axis internal/magnetic sensors package | Kalman-based fusion algorithm | 1. Accurate tracking is achieved by use of kalman filter to eliminate drift error. 2. Time lag is generated due to kalman filter. | 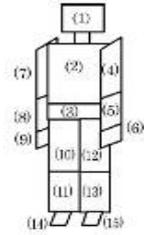 |
| Lee, Mun Wai, and Ramakant Nevatia [14] | Human pose tracking in monocular sequence using multilevel structured models | Grid-based belief propagation algorithm, data-driven Markov chain Monte Carlo | 1. Less position error due to full pose inference. 2. Longer processing time for rendering hence not suitable for real time application. | 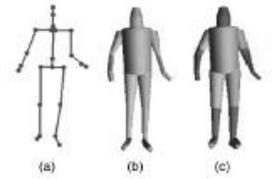 |
| Peursum, Patrick, Svetha Venkatesh, and Geoff West [15] | Smoothing for particle-filtered 3d human body tracking | Annealed Particle Filter, Particle Filter, factored-state hierarchical hidden Markov model | 1. Smoothed-inference techniques are implemented 2. Occlusion and poor segmentation is handled by hierarchical hidden markov model. 3. Tracking results are not so accurate. 4. Smoothing does not improve body tracking accuracy. 5. Processing time is increased due to smoothing. | 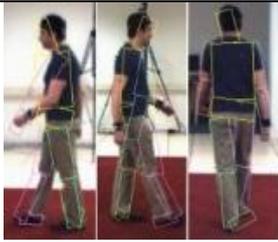 |

| Caillette, Fabrice, Aphrodite Galata, and Toby Howard [16] | Real-Time 3-D Human Body Tracking using Learnt Models of Behaviour | Variable Length Markov Models, monte-Carlo Bayesian frame-work | 1. Capable of handling fast and complex motions in real-time. 2. Body movements are captured while eliminating jitters. 3. Algorithm is robustness and efficiency 4. Simultaneously tracking of multiple subjects has not yet been fully investigated. 5. Dimensionality reduction is needed on learning cluster. | |
|---|---|---|---|---|

## 3. REVIEW OF DEPTH CAMERAS AND LIBRARIES SUPPORTED

In Table 3 presents a survey on depth cameras available in market with their specifications, and Table 4 gives a comparison of different Natural User Interface (NUI) libraries available with advantages and disadvantages.

Table 3. Survey of different depth cameras with their specifications.

| Specifications | Kinect Camera | Sony PlayStation Eye | Prime Sense Sensor | Intel's Creative Camera |
|---|---|---|---|---|
| Illustration | 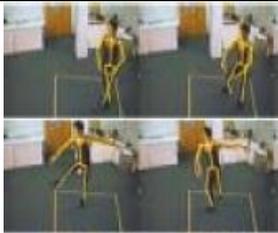 | | | |
| Viewing angle | 43°vertical by 57° horizontal. | 56° to 75° field of view. | 57.5°to 45° field of View | 73° field of view (diagonal). |
| Device range | Minimum 0.8 meter to maximum 4 meter. | Minimum 0.3meter | Minimum 0.8 meter to maximum 3.5 meter. | Minimum 0.15 meter to maximum 0.99 meter. |
| Frame rate | 12 and 30 frames per second (FPS). | 75and 187 frames per second (FPS). | 60 frames per second (FPS). | 30 frames per second (FPS). |
| Resolution | 1280 x 960 resolution at 12 frames per second, or a 640 x 480 resolution at 30 frames per second. | 320 x 240 resolution at 187 frames per second, or a 640 x 480 resolution at 75 frames per second. | 640 x 480 resolution. | 1280x720resolution. |
| IR camera | Yes. | No | Yes. | Yes. |
| Microphone array | Yes. | Yes. | Yes. | Yes. |
| OS support | Windows. | Windows, Mac OS, Linux. | Windows, Linux. | Windows. |

| Comments | 1. Widely used for gaming and application development. 2. Fastest selling electronic device. 3. Drivers are made available from Microsoft as well as third party drivers are also available. | 1. Drivers are not still available from Sony. 2. PlayStation playing experience is enhanced. | 1. Best depth performance. 2. Low-power consumption. 3. OpenNI compatible. | 1. Very limited device range. 2. Portable camera with HD support. 3. Drivers are made available from Intel. |
|---|---|---|---|---|

Table 4. A comparison table for Natural User Interface (NUI) libraries.

| # | Techniques | Pros | Cons |
|---|---|---|---|
| 1 | Microsoft Kinect SDK | 1. Easy to install, fairly widespread. 2. New version support skeleton tracking. 3. Does not require camera calibration. 4. Able to grab the full 1280 x 960 resolution of the camera. 5. Predictive tracking of joints. 6. Skeleton Recognition is done very fast. 7. Joints occlusion handled. 8. Description of the SDK architecture and documentation for the APIs. | 1. Support for windows only. 2. No skeleton tracking. 3. Limited language support, only for C/C++ and C#. 4. Higher processing power. |
| 2 | OpenNI/NITE | 1. Very popular, and ready to use methods. 2. Support skeleton tracking. 3. Available for most languages. 4. Any OS compatible. | 1. Difficult to install. 2. Calibration pose is required. 3. OpenNI is limited to 800x600 resolution. 4. No predictive tracking. 5. Joints occlusion not handled properly. 6. Gets confused with very fast movements. |
| 3 | Libfreenect | 1. Support for several applications. 2. Any OS compatible. 3. Available for most languages. | 1. Difficult to install. 2. No skeleton tracking. |
| 4 | CL NUI | 1. Can capture wide range of body movements. 2. Camera noise can be filtered. | 1. Cannot perform motion prediction. 2. No support for occlusion handling. |
| 5 | Evoluce SDK | 1. Support various gesture recognition methods. 2. Easy to install. 3. Support skeleton tracking. | 1. Only for Windows 7. 2. Calibration pose is required. 3. Limited language support, only for C/C++ and C#. |
| 6 | Delicode NImate | 1. Quite fast. 2. Support skeleton tracking. 3. Does not require camera calibration. | 1. Skeleton tracking not done properly. 2. Only for Windows. |

## 4. METHODOLOGY

From the above survey, it has been observed that detecting skeleton joints and tracking is a major problem. So kinect camera is a better option to get depth information of human body which is used in the proposed system.

Some researchers have tried to use more than one camera to detect and determine the depth of an object, which increases the cost and the system slows down due to increased data processing. Fortunately, due to the advancement in camera technology, depth camera such as kinect sensor makes it possible to get the depth of an object.

As using the depth information provided by kinect camera, each pixel corresponds to the estimate of the distance between the kinect camera and the closest object in the scene at that pixel's location, with the help of this information kinect camera allows to track different parts of human body in three dimension.

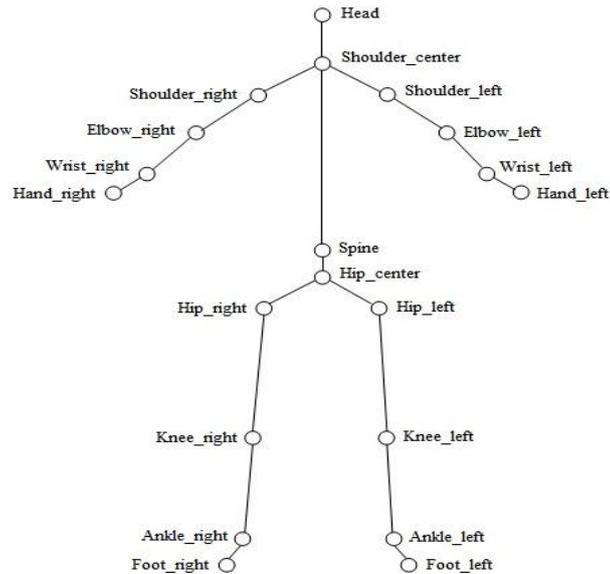

Figure 1. Kinect skeleton model of joints

Kinect camera allows us to produce depth, texture, user and skeleton information. The depth information is obtained from IR cameras on kinect; the texture information is the RGB color map of the scene which can be obtained through the RGB camera on the kinect. The user information is obtained from the binary images which includes the detected people in the scene. To obtain the skeleton information the person has to stand in front of kinect camera, as we are using MS kinect SDK calibration pose is not required and kinect tracks the human skeleton in real time. The MS kinect SDK is a middle ware framework supported by kinect camera; we get 20 joint positions using MS kinect SDK. The skeleton joint positions obtained from kinect camera are shown in figure 1. Following are the major steps involved in the proposed system.

Steps:

1. Skeleton Recognition and Tracking
2. 3D Human Model Creation
3. Rigging
4. Application of Motion Data to Rig

The Table 5 shows the dataset defined for the system to apply for sports; the gestures involved are recognised by the kinect camera. The following figure 2 shows the proposed methodology of our system, the system consist of the following major phases in implementation.

Table 5.  Anticipated dataset applied to sports.

| # | Gestures | Joints Involved | Additional Information |
|---|---|---|---|
| 1 | Sprinting | Hip, Knee, Ankle | Short running |
| 2 | Jumping | Hip, Knee, Ankle | Distance from ground to jumped object |
| 3 | 1 Hand wave | Hand, Wrist, Elbow, Shoulder | Left/Right hand wave |
| 4 | 2 Hands wave | Hand, Wrist, Elbow, Shoulder | Both hands wave |
| 5 | Throwing | Shoulder, Elbow, Wrist | Throwing a ball |
| 6 | Heading | Head, Shoulder | Heading a ball |
| 7 | Kicking | Hip, Knee, Ankle | Kicking a ball |

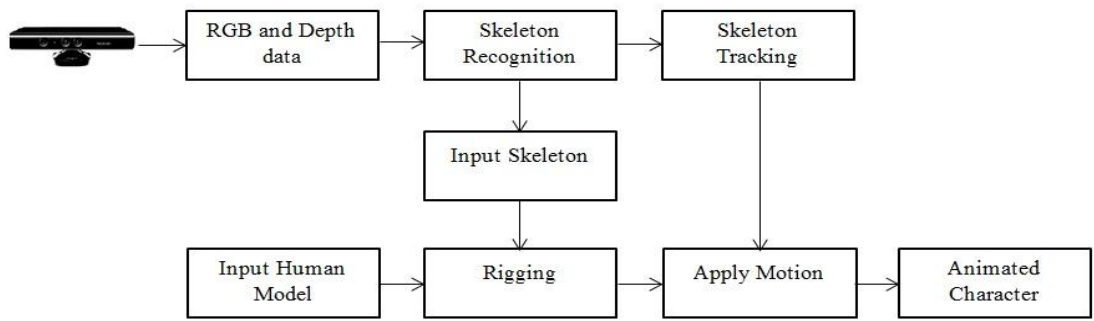

Figure 2.  System block diagram

### 4.1. Skeleton Recognition and Tracking

The process of extracting skeleton of the body from the input data is termed as skeleton recognition and capturing the movements of each joint position frame by frame is termed as skeleton tracking.

The skeleton recognition is done using Kinect camera, the process includes separating foreground from background; once the foreground is obtained, human body is separated out and the human body is segmented into different regions to get the skeleton joints. This technique is done using the algorithm based on the paper by Shotton and et. al. [8]. The skeleton tracking is done using MS Kinect SDK framework.

### 4.2. 3D Human Model Creation

The 3D human model is created using open source software of MakeHuman and student version of Autodesk Maya. The process consists of following phases.
1.      Mesh model is created.
2.      Texture is applied on the mesh.
3.      Clothing is applied on the human model.

### 4.3. Rigging

Rigging is the process of attaching skeleton to a human model; the human model is prepared using open source software of Make Human. The joints of the skeleton need to be placed at corresponding positions on the created human model in order to map rig character skeleton with captured kinect skeleton.

The input to rigging phase is the human model created and the input skeleton obtained from the kinect camera, to perform the rigging process we use the algorithm based on the paper by Ilya Baran and Jovan Popovic [22]. The output of the rigging phase is skeleton attached to the human model.

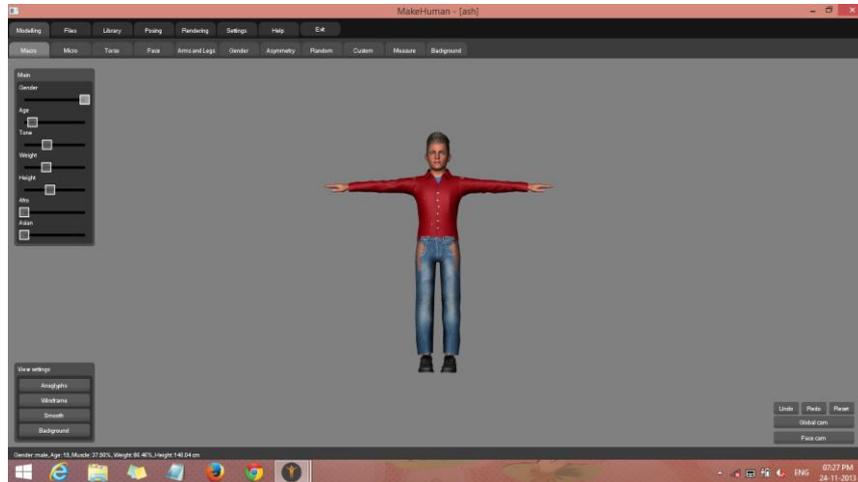

Figure 3. 3D human model created using MakeHuman.

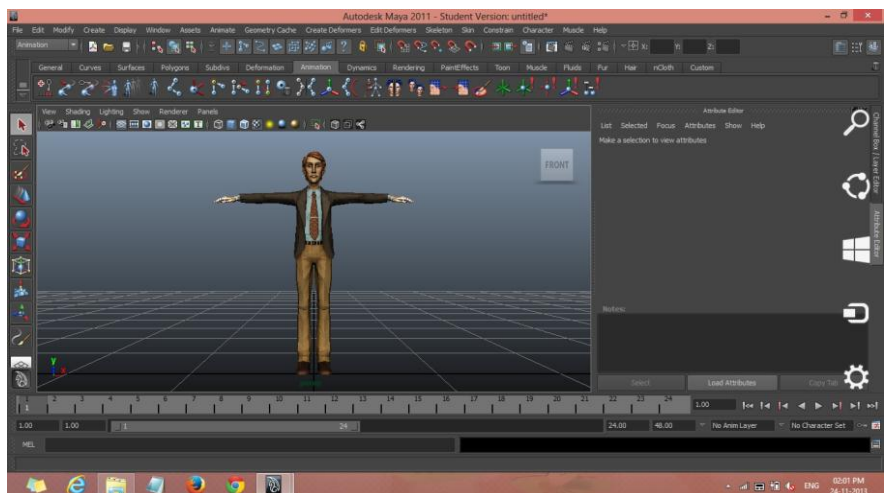

Figure 4. 3D human model created using Autodesk Maya.

## 4.4. Application of Motion Data to Rig

Once the rigging process is completed we need to apply the motion data that we get from skeleton tracking phase to rig character to perform animation. The data obtained from skeleton tracking phase is relevant only to the position of joints. Making the rig character to follow these positions is not correct; due to mismatch of size. Hence we need to extract rotations obtained from skeleton tracking phase to the rig character, to do so we use the concept of motion data transformation using rotation matrix.

To compute the rotation matrix, we need to get the rotation angles of all the joints relatively to x, y and z axis. But we have only the joints positions in every frame. In order to compute all joints rotation matrix A satisfying equation $Y = AX$, where Y, X are the transformed and original vertices respectively and A is the rotation matrix. To compute the original rotation matrix M (v, θ) through the following equation using a unit vector v = (x, y, z) which is perpendicular to a, b. Where a is vector of one bone of the last frame and b is vector of same bone in current frame.

$$M(v, \theta) = \begin{bmatrix} cos\theta + (1-cos\theta)x^2 & (1-cos\theta)xy - (sin\theta)z & (1-cos\theta)xz + (sin\theta)y \\ (1-cos\theta)yx + (sin\theta)z & cos\theta + (1-cos\theta)y^2 & (1-cos\theta)yz - (sin\theta)x \\ (1-cos\theta)zx - (sin\theta)y & (1-cos\theta)zy + (sin\theta)x & cos\theta + (1-cos\theta)z^2 \end{bmatrix} \quad (1)$$

The final rotation matrix is computed as:

$$R(j) = \prod_{i=1}^{n} M(v, \theta) \quad (2)$$

Once the rotations are transformed to the rig character, the rig character performs similar actions as that of the performer, which will be the output of the system.

## 5. MATHEMATICAL MODEL

The mathematical model for the proposed system is stated below.
Objective: Real time 3D character animation using markerless motion capture.
Let S be the system, such that
S = {I, O, F, Su, Fa}
I = Input to the system
O = Output of the system
F = Set of functions
Su = Success
Fa = Failure

1. Input
    I = {R, D}
    R = {r1, r2, r3….rn} - set of RGB image frames.
    D = {d1, d2, d3….dn} - set of depth image frames.
2. Output
    O = {animated character which performs same action as that of the performer}.
3. Functions
    F = {SR, ST, RI, AM}
    SR- Skeleton recognition
    SR = {j1, j2, j3…..j20} – set of skeleton joints.
    ST – Skeleton tracking
Let Y be the Transformed matrix such that
Y = AX

Where Y, X are the transformed and original vertices respectively and A is the rotation matrix.
Assumptions:

1. Assume the motion of character is coherent.
2. The current rotation of each joint is an addition of the rotations before.
Based on above assumptions we compute the rotation matrix M (v, θ), as shown in equation (1) using a single rotation angle θ and a unit vector v = (x, y, z) which is perpendicular to a, b.
The final rotation matrix is computed as shown in equation (2).
4. Success
    Su= {character performs similar action as that of the performer}.

5. Failure
    Fa = {1. Character does not perform any action
    2. Character performs dissimilar action}.

## 6. CONCLUSIONS

After conducting a survey on different motion capture and skeleton tracking technique, it is found that there is lot of scope for the development of such system. Hence, we proposed a system using markerless motion capture for 3D human character animation using kinect camera, which takes comparatively less development and processing time, this technique can widely be applied for gaming and film industry.

We have also done survey on various depth cameras available and different NUI libraries available for development with these cameras.